\documentclass[twocolumn,amsmath,amssymb,prl,aps,showkeys]{revtex4}
\usepackage{graphicx}
\usepackage{amsmath,amssymb}
\usepackage{mathrsfs}
\usepackage{color}
\usepackage{afterpage}
\usepackage[version=3]{mhchem}
\usepackage[normal]{subfigure}
\usepackage{natbib}
\usepackage{soul}
\usepackage{color}
\usepackage[normalem]{ulem}

\begin{document}
\title{Structure and electronic properties of transition-metal/Mg bimetallic clusters at realistic temperatures and oxygen partial pressures}
\author{Shikha Saini$^1$, Debalaya Sarker$^{1}$, Pooja Basera$^1$, Sergey V. Levchenko$^{2,3}$, Luca M. Ghiringhelli$^2$, Saswata Bhattacharya$^1$\footnote{saswata@physics.iitd.ac.in}} 
\affiliation{$^1$Dept. of Physics, Indian Institute of Technology Delhi, New Delhi 110016, India\\$^2$Theory Department, Fritz-Haber-Institut der Max-Planck-Gesellschaft, Faradayweg 4-6, D-14195 Berlin, Germany\\$^3$Laboratory of Modelling and Development of New Materials, NUST MISIS, Leninskiy prospekt 4, Moscow 119049, Russia}
\date{\today}
\begin{abstract}
Composition, atomic structure, and electronic properties of TM$_x$Mg$_y$O$_z$ clusters (TM = Cr, Ni, Fe, Co, $x+y \leq 3$) at realistic temperature $T$ and partial oxygen pressure $p_{\textrm{O}_2}$ conditions are explored using the {\em ab initio} atomistic thermodynamics approach. The low-energy isomers of the different clusters are identified using a massively parallel cascade genetic algorithm at the hybrid density-functional level of theory. On analyzing a large set of data, we find that the fundamental gap E$_\textrm{g}$ of the thermodynamically stable clusters are strongly affected by the presence of Mg-coordinated O$_2$ moieties. In contrast, the nature of the transition metal does not play a significant role in determining E$_\textrm{g}$. Using E$_\textrm{g}$ of a cluster as a descriptor of its redox properties, our finding is against the conventional belief that the transition metal plays the key role in determining the electronic and therefore chemical properties of the clusters. High reactivity may be correlated more strongly with oxygen content in the cluster than with any specific TM type.
\end{abstract}
\maketitle
\section{Introduction}
Transition-metal (TM) nanoparticles/clusters are used as catalysts for a variety of important chemical processes~\cite{r1,r2,r3,r5,r6}. At nanoscale, the particles acquire unique properties not found in the bulk TMs. The size of the nanoparticles is an additional parameter that affects their electronic and chemical properties and can be used to control the activity and selectivity of the catalyst. Moreover, the particles containing two or more transition or other metals can be synthesized, which tremendously increases the possibilities for tuning their functional properties. Together with these possibilities, new challenges arise for the design of more efficient and stable nano-catalysts. In particular, due to the increased complexity, the atomic structure and stability of the nanoparticles is harder to determine. Moreover, the particles usually operate in a reactive atmosphere, containing, e.g., oxygen. The interaction of the particles with oxygen will eventually result in their partial or complete oxidation, whether it is intended or not~\cite{bhattacharya2014efficient}. In fact, in some applications~\cite{r10}, the particles are created by a reduction of the oxide in hydrogen. Thus, both oxidized and reduced metal atoms are usually present at the surface, with their relative concentrations depending on the temperature and oxygen partial pressure. Disentangling the role of different oxidation states in a catalytic process is extremely challenging. 

In view of the above, it is important to provide theoretical guidance to experiment and technology on the composition and atomic structure of metal particles in a reactive atmosphere, in particular in the ubiquitous presence of oxygen. Stability being a key element for the desired functioning of a catalyst, it is a prerequisite to know the most stable phase of a particle/cluster in the reactive atmosphere. There are both experimental and theoretical reports stating the existence of particular phases viz. Ni$_4$O$_5$ clusters (on top of MgO substrate) under oxygen atmosphere~\cite{Smerieri2015}. However, no quantitative information has been provided so far to understand which structures/compositions are stable at what experimental conditions.

We report here an {\em ab initio} atomistic thermodynamics~\cite{scheffler1986} study of TM clusters supported on MgO, as well as unsupported TM-Mg clusters in an oxygen atmosphere. Gas-phase clusters are often used as model systems to study structural and compositional variability of the clusters at realistic conditions both experimentally and theoretically. Although such models cannot capture the full complexity of the supported nano-catalysts, they provide the necessary knowledge to disentangle various effects determining the catalytic activity. Using a massively parallel {\em ab initio} cascade genetic algorithm~\cite{bhattacharya2014efficient}, we identify stable and metastable structures of the clusters at a hybrid density functional theory (DFT) level, in order to overcome the difficulties of the standard Local Density Approximation (LDA) and Generalized Gradient Approximation (GGA) exchange-correlation functionals in describing localization of $d$-electrons in TMs, as well as the charge transfer from metal atoms to oxygen~\cite{bhattacharya2013stability}. We present an exhaustive and reliable database of small ternary TM$_x$Mg$_y$O$_z$ clusters (TM = Cr, Ni, Fe, Co; $x+y$ $\leq$ 3, $z$ = 1, 2, ...,12). Among a variety of TMs, Cr, Ni, Fe, and Co are of special interest to the catalytic research community because of their wide applications viz. in partial oxidation of CH$_4$~\cite{r7,r8,r9,r10,aiken1999review,hill1995homogeneous}, selective oxidation of alkanes~\cite{r22,r23,r24,r25}, alcohols~\cite{r26,r27}, olefins~\cite{r28}, and aromatics~\cite{r29}, selective reduction of nitrogen oxides~\cite{r30}, and oxidation of hydrogen sulfide~\cite{r31}. MgO is often used as a support in these applications, or as a component of an oxide solid solution catalyst~\cite{r10}. The high thermal stability and low cost are two important factors that make MgO to be one of the best choices for catalytic applications~\cite{Smerieri2015, Pal2014}.

A simple descriptor that captures both the nature of the reactive oxygen centers and the tendency of adjacent TM centers to gain electron density is the ligand-to-metal charge transfer excitation energy~\cite{r49}. For oxides having TM centres in their highest oxidation states, this ligand-to-metal charge transfer excitation energy typically corresponds to the band-gap energy. For a cluster, electron transfer to/from it is more appropriately described by the electron affinity (EA) and ionization potential (IP), respectively. Therefore, the clusters having a high EA and a low IP are the ones having a low fundamental gap ($E_\textrm{g} = $ EA - IP). Clusters possessing a high EA and/or a low IP may be expected to be better catalysts because they would accept or donate an electron more readily~\cite{SaswataPRBR2015, NoaPRL2012}. Thus, fundamental gap E$_\textrm{g}$ can be considered as an important feature of the clusters that determines their redox properties, and in many cases can serve as a descriptor of the catalytic activity~\cite{r43, r44, r49, r50, r47,r48, r45,r51}.

We therefore analyze the dependence of E$_\textrm{g}$ of the clusters on composition, temperature, and O$_2$ pressure, and determine the explicit role of different atomic species to modulate the E$_\textrm{g}$ values. Strikingly, we find that the nature of TM plays a negligible role in tuning the E$_\textrm{g}$ in TM$_x$Mg$_y$O$_z$ clusters. Rather, E$_\textrm{g}$ is strongly correlated with the presence of Mg-coordinated O$_2$ (i.e., the oxygen atoms directly attached with Mg atom) and is found to be minimum under O-rich conditions (i.e., z $>$ x+y, when $T$ $\approx$ 200K - 450K and $p_{\textrm{O}_2}$ $\approx$ 10$^{-6}$ - 10$^{10}$ atm). The analysis of our high-throughput data suggests that the tunability of the cluster chemical properties may be limited at given ($T$,$p_{\textrm{O}_2}$) conditions, but is strongly dependent on the conditions.

\section{Methodology}
We have considered first a wide range of TM$_x$Mg$_y$O$_z$ cluster compositions, where $z$ is determined by thermal equilibrium with the environment at given temperature $T$ and partial oxygen pressure $p_{\textrm O_2}$~\footnote{We varied $z$ value starting from 1, 2, 3,... and kept on increasing it until the specific TM$_x$Mg$_y$O$_z$ stoichiometry goes totally outside our phase diagrams, i.e. under no circumstances that stoichiometry can be important at realistic conditions}. In order to get the minimum energy configurations, for each stoichiometry, the total energy is minimized with respect to both geometry and spin state. The low-energy structures (including the global minimum) are generated from an exhaustive scanning of the potential energy surface~(PES) using our recent implementation of cascade genetic algorithm (GA)~\cite{bhattacharya2013stability,bhattacharya2014efficient}. The term ``cascade'' means a multi-stepped algorithm where successive steps employ higher level of theory and each of the next level takes information obtained at its immediate lower level. Typically here the cascade GA starts from a DFT with semi-local xc-functionals and goes up to DFT with hybrid xc-functionals. This GA algorithm's implementation is thoroughly benchmarked and it's efficiency is validated (w.r.t more advanced theory) in detail in Ref.~\cite{bhattacharya2014efficient}.

We have performed the DFT calculations using FHI-aims, which is an all electron code with numerical atom centred basis sets~\cite{blum2009ab}. The low-energy GA structures are further optimized at a higher level settings, where energy minimization is performed with vdW-corrected PBE (PBE+vdW) functional, ``tight - tier 2'' settings, and force tolerance was set to better than 10$^{-5}$ eV/${\textrm \AA}$.  The van der Waals correction is calculated as implemented in Tkatchenko-Scheffler scheme~\cite{tkatchenko2009accurate}. The total single-point energy is calculated afterwards on top of this optimized structure via vdW-corrected-PBE0~\cite{perdew1996rationale} hybrid xc-functional (PBE0+vdW), with ``tight - tier 2'' settings \footnote {As described in details in Ref~\cite{bhattacharya2014efficient, SaswataPRBR2015}, in our cascade GA, the latter energy is used to evaluate the fitness function, i.e., a mapping of the energy interval between highest-and lowest-energy cluster in the running pool into the interval [0, 1]. Obviously, the higher the value of the fitness function for a cluster, the higher is the probability of selecting it for generating a new structure.}. Note that PBE+vdW strongly overestimates the stability of oxide-clusters under O-rich conditions (i.e. with larger $z$ values) as reported for the case of Mg$_y$O$_z$ clusters~\cite{bhattacharya2013stability}. This gives a qualitatively wrong prediction that adsorption of O$_2$  could be favored over desorption up to a large excess of oxygen. Such behavior is not confirmed by hybrid functionals [e.g. HSE06, PBE0] employed in our calculations. For the data set we have used, the difference in energetics of PBE0 and HSE06 is always within 0.04~eV. The spin states of the clusters are also sometimes different as found by PBE and PBE0/HSE06. Thus, all our results are thoroughly tested and benchmarked w.r.t hybrid xc-functionals (PBE0) using ``tight'' numerical settings and tier 2 basis set. For estimation of (vertical) electron affinity (VEA), (vertical) ionization potential (VIP), and fundamental gap (E$_\textrm{g}$ = VEA - VIP) we have used the $G_0W_0$@PBE0 approach with ``really-tight'' numerical settings and tier 4 basis set~\cite{blum2009ab}.

The free energy of the isomers \footnote{We considered all the isomers within an energy window of 0.5~eV from the global minimum as we have seen that it's very unlikely that isomers above 0.5~eV from the GM would become more stable after including their translational, rotational, vibrational, spin and symmetry free energy contributions to the total energy.} is then calculated as a function of $T$ and $p_{\textrm O_2}$ for each stoichiometry using the {\em ab initio} atomistic thermodynamics (aiAT) approach. The concept of aiAT was successfully applied initially for bulk semiconductors~\cite{scheffler1986resonant, scheffler1988parameter}, and later applied to the study of oxide formation at the surface of some TMs and other materials~\cite{wang1998hematite,lee2000gaas,reuter2003composition,reuter2005ab}. We have recently extended this approach to clusters in a reactive atmosphere~\cite{bhattacharya2013stability,bhattacharya2014efficient,sasprmr17, sasjpcl}. From different cluster compositions and structures with the lowest free energy the thermodynamic phase diagram can be constructed as a function of $T$ and $p_{\textrm O_2}$. Such phase diagrams are shown in Fig.~\ref{fig2}. At a given $T$, $p_{{\textrm O}_2}$, and $x, y$, the stable stoichiometry of a TM$_x$Mg$_y$O$_z$ cluster is determined via aiAT, i.e., by minimizing the Gibbs free energy of formation $\Delta G_f (T,p_{{\textrm O}_2})$.
\begin{equation}
\begin{split}
\Delta G_f (T,p_{{\textrm O}_2}) = F_{{\textrm {TM}}_x{\textrm {Mg}}_y{\textrm O}_z}(T) -  F_{{\textrm {TM}}_x{\textrm {Mg}}_y}(T) \\ - z\times\mu_\textrm{O}(T,p_{{\textrm O}_2})
\end{split}
\label{eq1}
\end{equation}
Here, $F_{{\textrm {TM}}_x{\textrm {Mg}}_y{\textrm O}_z}(T)$ and $F_{{\textrm {TM}}_x{\textrm {Mg}}_y}(T)$ are the Helmholtz free energies of the ${\textrm {TM}}_x{\textrm {Mg}}_y{\textrm O}_z$ and the pristine ${\textrm {TM}}_x{\textrm {Mg}}_y$ cluster~\footnote {The clusters are at their ground state configuration w. r. t. geometry and spin respectively.} and $\mu_\textrm{O}(T,p_{{\textrm O}_2})$ is the chemical potential of oxygen. As explained in Ref.~\cite{bhattacharya2014efficient}, $F_{{\textrm {TM}}_x{\textrm {Mg}}_y{\textrm O}_z}(T)$ and $F_{{\textrm {TM}}_x{\textrm {Mg}}_y}(T)$ are calculated as the sum of DFT total energy, DFT vibrational free energy (up to harmonic approximation), translational, rotational, symmetry and spin-degeneracy free-energy contributions. The dependence of $\mu_\textrm{O}(T,p_{{\textrm O}_2})$ on $T$ and $p_{{\textrm O}_2}$ is calculated using the ideal (diatomic) gas approximation with the same DFT functional as for the clusters. The phase diagram for a particular ${\textrm {TM}}_x{\textrm {Mg}}_y{\textrm O}_z$ is constructed by identifying the lowest free-energy structures at each $T$, $p_{\textrm O_2}$. 

\section{Results and Discussions}
\begin{figure}[t]
	\centering
	\includegraphics[scale=0.4]{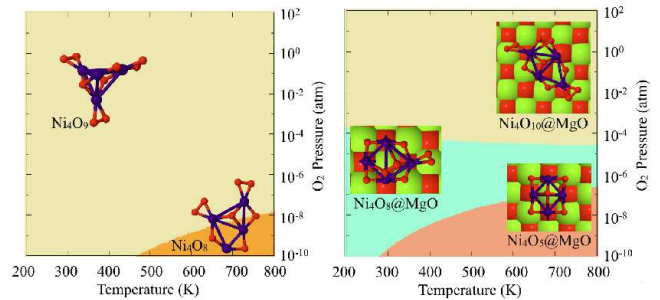}
	\caption{$T$-$p_{\textrm{O}_2}$ phase diagram (see text) of Ni$_4$ clusters under oxygen atmosphere with (right) and without (left) MgO substrate. The red, navyblue, and green color-balls represent O, Ni and Mg respectively.}
	\label{fig1}
\end{figure}

In order to understand whether the presence of MgO substrate is crucial or not in the stability analysis of TM clusters, we have considered two cases: (a) stand-alone Ni$_4$ cluster in an oxygen environment, and (b) Ni$_4$ cluster on MgO substrate in oxygen environment. It is indeed evident from Fig.~\ref{fig1} that ignoring the substrate's effect in stability analysis of the TM clusters is not meaningful as the substrate (here MgO) plays a notable role to the stability of the TM clusters under oxygen atmosphere. The corresponding $T$-$p_{\textrm{O}_2}$ phase diagrams are plotted in Fig.~\ref{fig1}  by minimizing the Gibbs' free energy of formation of the all possible configurations (see methodology for details)~\footnote{This phase diagram is plotted at the PBE+vdW level in order to show that the substrate plays crucial role in stability of configurations. The effect of advanced and accurate xc-functionals are introduced latter, while concluding our main results}. Note that in Ref.~\cite{Smerieri2015} the experimental evidence of Ni$_4$O$_5$ phase is observed from STM analysis. This phase is stable at a given experimental condition in the phase diagram only when we have considered the MgO substrate, while the concerned phase is totally missing when substrate effect is not included in our calculations. This clearly suggests that even if consideration of the support is computationally demanding, it is still a necessary condition to understand the stability of the TM clusters correctly. 

\begin{figure}[b!]
	\includegraphics[width=1.0\columnwidth,clip]{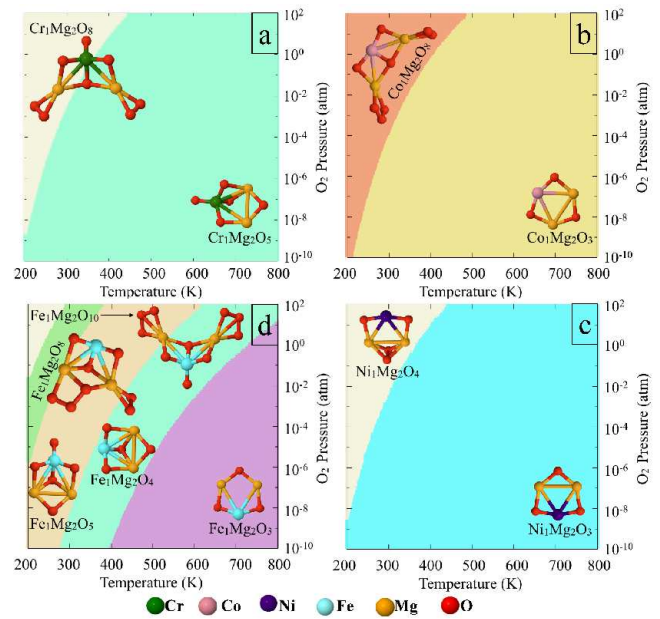}
	\caption{(color online) The most stable TM$_x$Mg$_y$O$_z$ clusters at various temperatures and pressures under thermodynamic equilibrium. In TM$_x$Mg$_y$O$_z$ clusters, TM = Cr (a), Co (b), Ni (c) and Fe (d). The geometries are optimized with PBE+vdW, and the electronic energy is calculated using HSE06+vdW. The vibrational free energy is computed under harmonic approximations.}
	\label{fig2}
\end{figure}

In order to explore the explicit effect of the support for a wide range of compositions and configurations, here we model the TM-MgO system by means of small bi-metallic TM$_x$Mg$_y$O$_z$ clusters.
An exhaustive set of TM$_x$Mg$_y$O$_z$ clusters are generated including all the possible structural and composition motifs, oxidations states, electronic spin, symmetry etc. as if when MgO-supported TM clusters are prepared from a solid solution. Following this thermodynamic phase diagrams are calculated for all possible combinations of $x+y \le 3$ ($y\neq 0$) in TM$_x$Mg$_y$O$_z$ clusters, and the most stable compositions and configurations are identified. Prior to this, all the global minimum structures of the clusters are determined by cascade GA. In Fig.~\ref{fig2}(a-d) we show phase diagrams of a set of TM$_1$Mg$_2$O$_z$ clusters, where four different TMs viz. Cr, Co, Ni, Fe are investigated respectively. The most stable configurations and the respective stable phases are shown within a window of experimentally achievable environmental conditions ($T$, $p_{\textrm O_2}$ window). From Fig.~\ref{fig2}a, we see for TM = Cr and $x$ = 1, $y$ = 2 at low $T$ and high $p_{\textrm O_2}$, Cr$_1$Mg$_2$O$_8$ is the most stable phase, while at high $T$ and low $p_{\textrm O_2}$  Cr$_1$Mg$_2$O$_5$ is the most stable phase. For TM = Co and Ni (see Fig.~\ref{fig2}b and Fig.~\ref{fig2}c) these are respectively (Co$_1$Mg$_2$O$_8$, Co$_1$Mg$_2$O$_3$), (Ni$_1$Mg$_2$O$_4$, Ni$_1$Mg$_2$O$_3$). However for TM = Fe (see Fig.~\ref{fig2}d) at low $T$ and high $p_{\textrm O_2}$ there are a lot of competing isomers viz. Fe$_1$Mg$_2$O$_{10}$, Fe$_1$Mg$_2$O$_8$ and Fe$_1$Mg$_2$O$_5$. At a moderate $T$ and moderate $p_{\textrm O_2}$ i.e. $T$ $\approx$300K, $p_{\textrm O_2}$ $\approx$ 10$^{-6}$ atm there is competition between two phases viz. Fe$_1$Mg$_2$O$_4$ and Fe$_1$Mg$_2$O$_5$. At high $T$ and low $p_{\textrm O_2}$ the most stable phase is Fe$_1$Mg$_2$O$_3$. Thus, we observe a general trend that usually O-rich clusters are more stable at low $T$ and moderate to high $p_{\textrm O_2}$ (i.e. upto $\approx$ 10$^{10}$ atm). 
\begin{figure}[t!]
\includegraphics[width=0.8\columnwidth,clip]{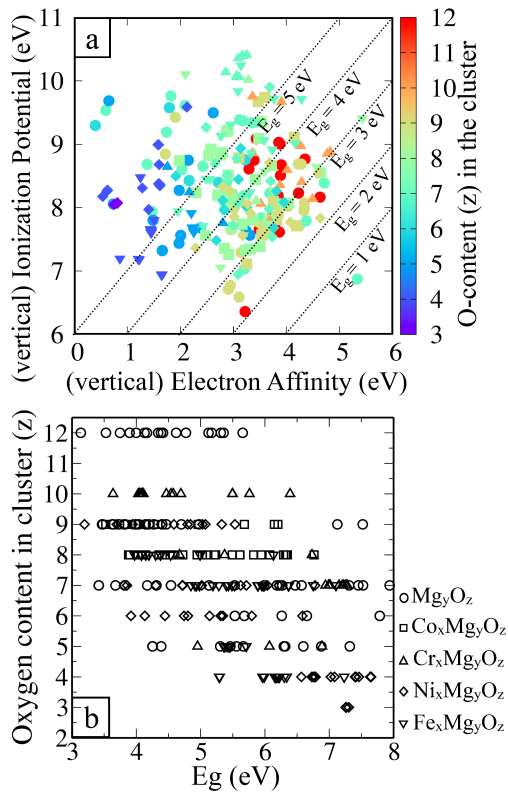}
\caption{(color online)  (a) VIP vs VEA for the low energy isomers stable at an experimentally achievable environmental condition for all different cluster sizes ($x+y\le 3$). The symbols represent the nature of the TM atoms in the clusters, while the colour is set as per the amount of oxygen content (as shown by the colour bar) in the cluster. The loci of constant E$_\textrm{g}$ are indicated by diagonal lines. (b) E$_\textrm{g}$ vs oxygen content ($z$) is plotted for all the stable isomers at realistic conditions (see text for details).}
\label{fig3}
\end{figure} 

The electronic structure of the clusters determines their reactivity, and can serve as a descriptor of the catalytic activity. Under reaction conditions, the catalyst comprises of a wide range of structures including different number of atoms with various oxidation states, all of which could be active to some extent in the catalytic reaction. Therefore, after identifying the most stable compositions (i.e., specific $x$, $y$, and $z$ in stable phases of TM$_x$Mg$_y$O$_z$), we have studied electronic structure of not only the global minimum isomer of that given composition but also all the low-energy isomers  lying within an energy window of 0.5~eV from the global minimum~\footnote{The energy window of 0.5~eV in total energy is carefully chosen such that within this error bar in energetics, the phase diagram will not change significantly in terms of stability of different phases.}.

In Fig.~\ref{fig3}a we show VEA and VIP values (at the level of $G_0W_0$@PBE0) of all such thermodynamically stable isomers. Typically we have included data for TM$_1$Mg$_1$O$_z$, TM$_1$Mg$_2$O$_z$, TM$_2$Mg$_1$O$_z$ with TM = Cr, Co, Ni, Fe. We have also included data for clusters without TM atoms, i.e., Mg$_y$O$_z$ ($y$ = 1-3) clusters. The symbols are selected as per the type of the TM atoms in the clusters, while the colour of the symbols are selected based on the amount of oxygen content in the cluster. The diagonal lines are drawn to represent the corresponding constant E$_\textrm{g}$, which is evaluated as the absolute difference between VIP and VEA. In Fig.~\ref{fig3}a the E$_\textrm{g}$ for the clusters are widely scattered at all different sizes and nature of TM atoms. In Fig.~\ref{fig3}b we have explicitly shown E$_\textrm{g}$ with different no. of oxygen content ($z$) in the clusters. Clearly, it is not following any trends except the E$_\textrm{g}$ for O-deficient clusters (i.e. $z\le x+y$) are consistently in the higher side (see, bluish points in the Fig.~\ref{fig3}a and the region around low oxygen content ($z \approx$ 3-5)). Thus for sure it can be concluded from Fig.~\ref{fig3} that O-deficient clusters (mostly stable at high $T$ and low $p_{\textrm O_2}$) are not good for catalysis. However, for O-rich clusters (i.e. $z\ge x+y$) both high and low E$_\textrm{g}$ are found (see Fig.~\ref{fig3}b around $z\ge 6$). This observation hints towards the fact that some structural feature in these O-rich clusters might act as active site that dominates the low  E$_\textrm{g}$ value. Moreover, the invariance of the  E$_\textrm{g}$ values for different TM clusters with same O-content emphasizes that it is not the nature of TM atoms but probably some active centers originating from oxygen moieties that affect the fundamental gap. Note that it is common understanding that the former (i.e., nature of the TM atoms) is solely responsible for making the material a good catalyst~\cite{hill1995homogeneous, aiken1999review, r7, r8, r9, r10, aiken1999review, hill1995homogeneous}.

We have next divided our clusters into two sets viz. (i) having low gap ($\le$ 4~eV) and (ii) having high gap ($>$ 4~eV). Following this, the vibrational frequencies  (corresponding O-vibrations) of all the structures are carefully analyzed.  
\begin{figure}[h!]
\includegraphics[width=0.8\columnwidth,clip]{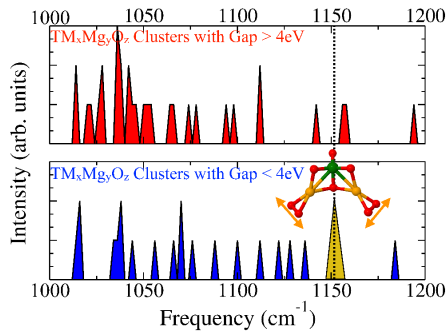}
\caption{(color online)  Histogram of vibrational frequencies ranging from 1000$-$1200~cm$^{-1}$ for clusters (i) gap $>$ 4~eV (top panel) (ii) gap $\le$ 4 eV (bottom panel).}
\label{fig4}
\end{figure} 
In Fig.~\ref{fig4} we show the histogram of vibrational frequencies by including an average of all vibrational spectra ranging 1000$-$1200~cm$^{-1}$. Clearly the peak corresponding to the frequency 1151~cm$^{-1}$ of the O$_2$ moiety bonded to the Mg atom, is missing for all clusters with the E$_\textrm{g}$ $>$~4 eV (see in Fig.~\ref{fig4}). This means clusters with low E$_\textrm{g}$ have some correlation with this Mg coordinated O$_2$ moieties, while clusters with higher E$_\textrm{g}$ do not have this structural feature.
\begin{figure}[t!]
\includegraphics[width=1.05\columnwidth,clip]{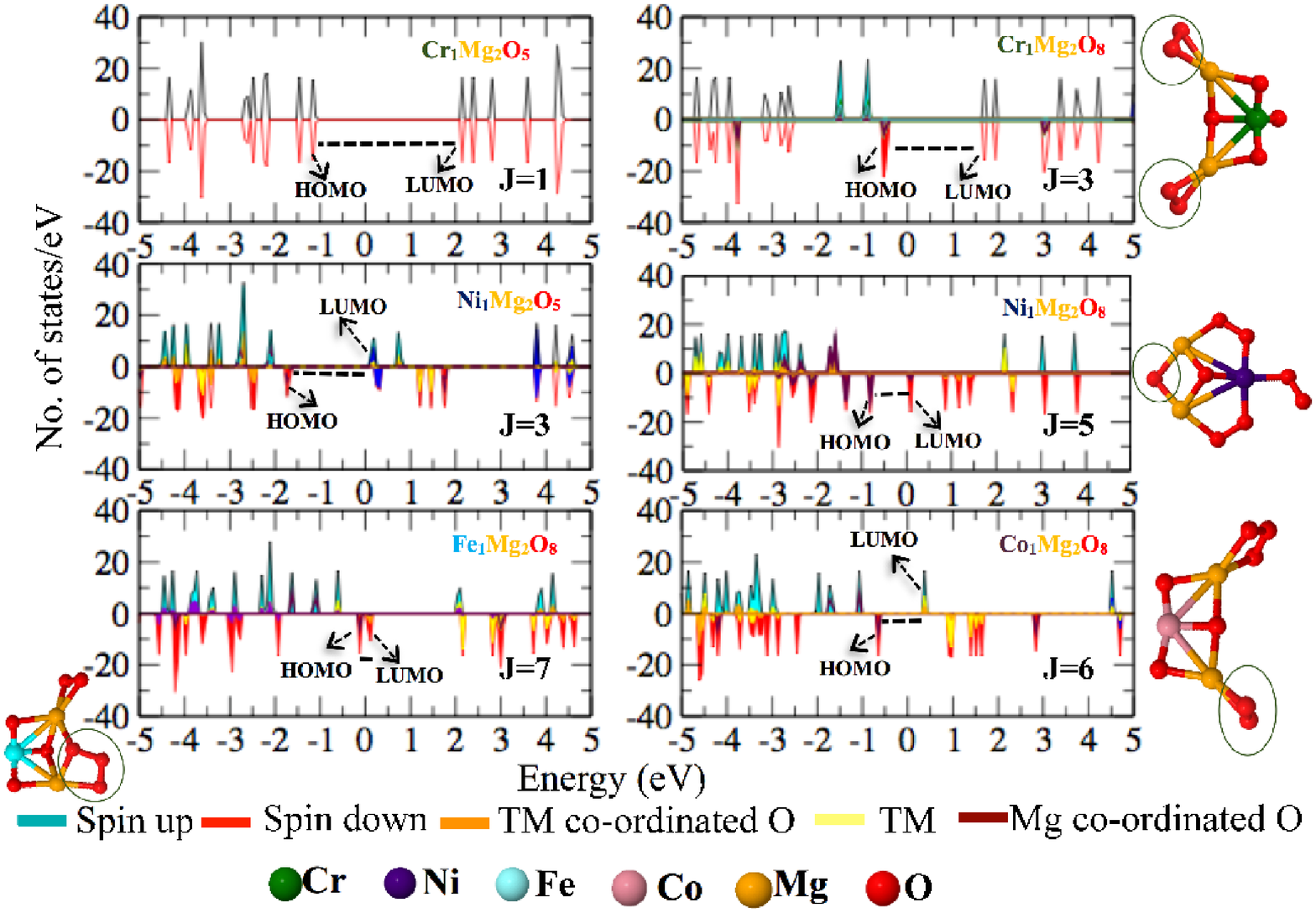}
\caption{(color online)  Atom-projected spin polarised density of states of different TM$_x$Mg$_y$O$_z$ clusters. The corresponding {\em spin multiplicity} (J) is also given in the respective plots.}
\label{fig5}
\end{figure} 

To unravel the particular role of these O$_2$ moieties in the E$_\textrm{g}$, we examined the density of states (DOS) profiles for different TM$_x$Mg$_y$O$_z$ clusters in Fig.~\ref{fig5}. We can see the HOMO-LUMO gap in Cr$_1$Mg$_2$O$_8$ is smaller than in Cr$_1$Mg$_2$O$_5$. Presence of spin-polarised states near fermi-energy (E$_f$) leads to the reduced HOMO-LUMO gap in the oxygen-rich phase. Further, comparing top panel in Fig.~\ref{fig5} we observe that the presence of TM (Cr) atom has minimal effect on the spin polarized states near E$_f$. These states occur only due to the extra oxygen atoms in Cr$_1$Mg$_2$O$_8$, whereas they are absent in case of Cr$_1$Mg$_2$O$_5$. To further understand if our observation is coincidental for Cr clusters or not, we have next plotted the same for Ni clusters in Fig.~\ref{fig5} (middle panel). We see here again that the HOMO-LUMO gap is smaller for the oxygen-rich phase, i.e., Ni$_1$Mg$_2$O$_8$ in comparison to Ni$_1$Mg$_2$O$_5$. 
By having the insight from the atom-projected DOS, we conclude that the spin-polarized states are indeed the states of Mg-coordinated oxygen atoms in the O-rich clusters, which have caused a reduced E$_\textrm{g}$ in these clusters. Next in Fig.~\ref{fig5} (bottom panel), we see the O-rich phases of Fe and Co clusters respectively, which are having smaller E$_\textrm{g}$ in comparison to their O-reduced phases. Again the atom-projected DOS indicate that the Mg coordinated O-atoms are playing pivotal role in the states near E$_f$ and hence are responsible for the reduced E$_\textrm{g}$ in these clusters. The Fe$_1$Mg$_2$O$_8$ cluster has the lowest E$_\textrm{g}$ among all because of the presence of spin-polarized 3$d$ states of Fe near E$_f$. However, it can easily be seen from the comparison of $z$ = 5 and $z$ = 8 clusters that it is not the nature of TM atoms, but the states of Mg coordinated O atoms in the O-rich phases of each TM$_x$Mg$_y$O$_z$ cluster that play the crucial role in controlling the E$_\textrm{g}$. Irrespective to any specific TM atoms, TM has always states at the LUMO level, whereas the HOMO keeps on shifting depending on oxygen content in the cluster. Under O-rich condition, the clusters usually have O$_2$ (containing loosely bound electrons) coordinated with Mg atom. This gives rise to occupied states near E$_f$ yielding a smaller E$_\textrm{g}$.
\section{Conclusion}
In summary, we have used a robust first-principles approach to understand the stability and electronic structure of ternary TM$_x$Mg$_y$O$_z$ clusters at realistic temperatures and oxygen partial pressures. The low-energy isomers of different composition are determined by cascade genetic algorithm. Following that, the stable compositions are found by minimizing Gibbs free energy of formation using aiAT methodology (within harmonic approximation for the vibrational contributions). From the analysis of the electronic structure of the clusters, we establish that the presence of Mg-coordinated-O in O-rich clusters (i.e. $z$ $>$ $x+y$) correlates with low E$_\textrm{g}$, whereas the nature of TM atoms is insignificant. We further note that these Mg coordinated O$_2$ moieties have unpaired electrons that give rise to multiple degenerate spin states. The latter play key role in producing a reduced E$_\textrm{g}$ from the O$_2$ moieties at an environmental condition $T$ $\approx$ 200K - 450K and $p_{\textrm{O}_2}$ $\approx$ 10$^{-6}$ - 10$^{10}$ atm; while at higher temperature and lower pressure stable clusters have less O-content. These observations lead us to argue that the chemical (redox) properties of mixed metal oxide clusters are sensitive to the oxygen content controlled by $T$ and $p_{\textrm{O}_2}$ rather than to the specific transition metal type. We hope that this fining will be helpful for understanding experiments with gas-phase and supported metal clusters at realistic ($T$, $p_{\textrm{O}_2}$) conditions, and for designing new functional nano-materials. 
\section{Acknowledgement}
SS acknowledges CSIR, India, for the senior research fellowship [grant no. 09/086(1231)2015-EMR-I]. PB acknowledges UGC, India, for the junior research fellowship [grant no. 20/12/2015(ii)EU-V]. SB and DS acknowledge the financial support from YSS-SERB research grant, DST, India (grant no. YSS/2015/001209). SB thanks Bryan R. Goldsmith for critically reading the manuscript. SVL is grateful for his support from the German Science Foundation in the frame of the Berlin Cluster of Excellence ``Unifying Concepts in Catalysis'', and to the Ministry of Education and Science of the Russian Federation in the framework of Increase Competitiveness Program of NUST MISIS (No K2-2016-013) implemented by a governmental decree dated 16 March 2013, No 211. We acknowledge the High Performance Computing (HPC) facility at IIT Delhi for computational resources. We thank Karsten Reuter for many helpful discussions. 


\begin{thebibliography}{47}
\expandafter\ifx\csname natexlab\endcsname\relax\def\natexlab#1{#1}\fi
\expandafter\ifx\csname bibnamefont\endcsname\relax
  \def\bibnamefont#1{#1}\fi
\expandafter\ifx\csname bibfnamefont\endcsname\relax
  \def\bibfnamefont#1{#1}\fi
\expandafter\ifx\csname citenamefont\endcsname\relax
  \def\citenamefont#1{#1}\fi
\expandafter\ifx\csname url\endcsname\relax
  \def\url#1{\texttt{#1}}\fi
\expandafter\ifx\csname urlprefix\endcsname\relax\def\urlprefix{URL }\fi
\providecommand{\bibinfo}[2]{#2}
\providecommand{\eprint}[2][]{\url{#2}}

\bibitem[{\citenamefont{McFarland and Metiu}(2013)}]{r1}
\bibinfo{author}{\bibfnamefont{E.~W.} \bibnamefont{McFarland}}
  \bibnamefont{and} \bibinfo{author}{\bibfnamefont{H.}~\bibnamefont{Metiu}},
  \bibinfo{journal}{Chemical reviews} \textbf{\bibinfo{volume}{113}},
  \bibinfo{pages}{4391} (\bibinfo{year}{2013}).

\bibitem[{\citenamefont{Zemski et~al.}(2002)\citenamefont{Zemski, Justes, and
  Castleman}}]{r2}
\bibinfo{author}{\bibfnamefont{K.}~\bibnamefont{Zemski}},
  \bibinfo{author}{\bibfnamefont{D.}~\bibnamefont{Justes}}, \bibnamefont{and}
  \bibinfo{author}{\bibfnamefont{A.}~\bibnamefont{Castleman}},
  \emph{\bibinfo{title}{Studies of metal oxide clusters: Elucidating reactive
  sites responsible for the activity of transition metal oxide catalysts}}
  (\bibinfo{year}{2002}).

\bibitem[{\citenamefont{Guzman and Gates}(2003)}]{r3}
\bibinfo{author}{\bibfnamefont{J.}~\bibnamefont{Guzman}} \bibnamefont{and}
  \bibinfo{author}{\bibfnamefont{B.~C.} \bibnamefont{Gates}},
  \bibinfo{journal}{Dalton Transactions} pp. \bibinfo{pages}{3303--3318}
  (\bibinfo{year}{2003}).

\bibitem[{\citenamefont{Liu et~al.}(2013)\citenamefont{Liu, Ji, Zou, Gu, Deng,
  Gao, Tang, and Dong}}]{r5}
\bibinfo{author}{\bibfnamefont{L.}~\bibnamefont{Liu}},
  \bibinfo{author}{\bibfnamefont{Z.}~\bibnamefont{Ji}},
  \bibinfo{author}{\bibfnamefont{W.}~\bibnamefont{Zou}},
  \bibinfo{author}{\bibfnamefont{X.}~\bibnamefont{Gu}},
  \bibinfo{author}{\bibfnamefont{Y.}~\bibnamefont{Deng}},
  \bibinfo{author}{\bibfnamefont{F.}~\bibnamefont{Gao}},
  \bibinfo{author}{\bibfnamefont{C.}~\bibnamefont{Tang}}, \bibnamefont{and}
  \bibinfo{author}{\bibfnamefont{L.}~\bibnamefont{Dong}}, \bibinfo{journal}{Acs
  Catalysis} \textbf{\bibinfo{volume}{3}}, \bibinfo{pages}{2052}
  (\bibinfo{year}{2013}).

\bibitem[{\citenamefont{Zhao et~al.}(2017)\citenamefont{Zhao, Yang, Chen, Liu,
  Ji, Zhang, Niu, Mao, Bao, Hu et~al.}}]{r6}
\bibinfo{author}{\bibfnamefont{G.}~\bibnamefont{Zhao}},
  \bibinfo{author}{\bibfnamefont{F.}~\bibnamefont{Yang}},
  \bibinfo{author}{\bibfnamefont{Z.}~\bibnamefont{Chen}},
  \bibinfo{author}{\bibfnamefont{Q.}~\bibnamefont{Liu}},
  \bibinfo{author}{\bibfnamefont{Y.}~\bibnamefont{Ji}},
  \bibinfo{author}{\bibfnamefont{Y.}~\bibnamefont{Zhang}},
  \bibinfo{author}{\bibfnamefont{Z.}~\bibnamefont{Niu}},
  \bibinfo{author}{\bibfnamefont{J.}~\bibnamefont{Mao}},
  \bibinfo{author}{\bibfnamefont{X.}~\bibnamefont{Bao}},
  \bibinfo{author}{\bibfnamefont{P.}~\bibnamefont{Hu}}, \bibnamefont{et~al.},
  \bibinfo{journal}{Nature communications} \textbf{\bibinfo{volume}{8}},
  \bibinfo{pages}{14039} (\bibinfo{year}{2017}).

\bibitem[{\citenamefont{Bhattacharya et~al.}(2014)\citenamefont{Bhattacharya,
  Levchenko, Ghiringhelli, and Scheffler}}]{bhattacharya2014efficient}
\bibinfo{author}{\bibfnamefont{S.}~\bibnamefont{Bhattacharya}},
  \bibinfo{author}{\bibfnamefont{S.~V.} \bibnamefont{Levchenko}},
  \bibinfo{author}{\bibfnamefont{L.~M.} \bibnamefont{Ghiringhelli}},
  \bibnamefont{and}
  \bibinfo{author}{\bibfnamefont{M.}~\bibnamefont{Scheffler}},
  \bibinfo{journal}{New Journal of Physics} \textbf{\bibinfo{volume}{16}},
  \bibinfo{pages}{123016} (\bibinfo{year}{2014}).

\bibitem[{\citenamefont{Hu and Ruckenstein}(2002)}]{r10}
\bibinfo{author}{\bibfnamefont{Y.~H.} \bibnamefont{Hu}} \bibnamefont{and}
  \bibinfo{author}{\bibfnamefont{E.}~\bibnamefont{Ruckenstein}},
  \bibinfo{journal}{Catalysis Reviews} \textbf{\bibinfo{volume}{44}},
  \bibinfo{pages}{423} (\bibinfo{year}{2002}).

\bibitem[{\citenamefont{Smerieri et~al.}(2015)\citenamefont{Smerieri, Pal,
  Savio, Vattuone, Ferrando, Tosoni, Giordano, Pacchioni, and
  Rocca}}]{Smerieri2015}
\bibinfo{author}{\bibfnamefont{M.}~\bibnamefont{Smerieri}},
  \bibinfo{author}{\bibfnamefont{J.}~\bibnamefont{Pal}},
  \bibinfo{author}{\bibfnamefont{L.}~\bibnamefont{Savio}},
  \bibinfo{author}{\bibfnamefont{L.}~\bibnamefont{Vattuone}},
  \bibinfo{author}{\bibfnamefont{R.}~\bibnamefont{Ferrando}},
  \bibinfo{author}{\bibfnamefont{S.}~\bibnamefont{Tosoni}},
  \bibinfo{author}{\bibfnamefont{L.}~\bibnamefont{Giordano}},
  \bibinfo{author}{\bibfnamefont{G.}~\bibnamefont{Pacchioni}},
  \bibnamefont{and} \bibinfo{author}{\bibfnamefont{M.}~\bibnamefont{Rocca}},
  \bibinfo{journal}{The Journal of Physical Chemistry Letters}
  \textbf{\bibinfo{volume}{6}}, \bibinfo{pages}{3104} (\bibinfo{year}{2015}),
  \bibinfo{note}{pMID: 26267209},
  \eprint{http://dx.doi.org/10.1021/acs.jpclett.5b01362},
  \urlprefix\url{http://dx.doi.org/10.1021/acs.jpclett.5b01362}.

\bibitem[{\citenamefont{Scheffler and Weinert}(1986)}]{scheffler1986}
\bibinfo{author}{\bibfnamefont{M.}~\bibnamefont{Scheffler}} \bibnamefont{and}
  \bibinfo{author}{\bibfnamefont{C.}~\bibnamefont{Weinert}}, in
  \emph{\bibinfo{booktitle}{Defects in Semiconductors}}, edited by
  \bibinfo{editor}{\bibfnamefont{H.~J.~v.} \bibnamefont{Bardeleben}}
  (\bibinfo{publisher}{Trans. Tech. Publ. Ltd, Switzerland},
  \bibinfo{year}{1986}), pp. \bibinfo{pages}{25--30}.

\bibitem[{\citenamefont{Bhattacharya et~al.}(2013)\citenamefont{Bhattacharya,
  Levchenko, Ghiringhelli, and Scheffler}}]{bhattacharya2013stability}
\bibinfo{author}{\bibfnamefont{S.}~\bibnamefont{Bhattacharya}},
  \bibinfo{author}{\bibfnamefont{S.~V.} \bibnamefont{Levchenko}},
  \bibinfo{author}{\bibfnamefont{L.~M.} \bibnamefont{Ghiringhelli}},
  \bibnamefont{and}
  \bibinfo{author}{\bibfnamefont{M.}~\bibnamefont{Scheffler}},
  \bibinfo{journal}{Physical review letters} \textbf{\bibinfo{volume}{111}},
  \bibinfo{pages}{135501} (\bibinfo{year}{2013}).

\bibitem[{\citenamefont{Subbaraman et~al.}(2012)\citenamefont{Subbaraman,
  Tripkovic, Chang, Strmcnik, Paulikas, Hirunsit, Chan, Greeley, Stamenkovic,
  and Markovic}}]{r7}
\bibinfo{author}{\bibfnamefont{R.}~\bibnamefont{Subbaraman}},
  \bibinfo{author}{\bibfnamefont{D.}~\bibnamefont{Tripkovic}},
  \bibinfo{author}{\bibfnamefont{K.-C.} \bibnamefont{Chang}},
  \bibinfo{author}{\bibfnamefont{D.}~\bibnamefont{Strmcnik}},
  \bibinfo{author}{\bibfnamefont{A.~P.} \bibnamefont{Paulikas}},
  \bibinfo{author}{\bibfnamefont{P.}~\bibnamefont{Hirunsit}},
  \bibinfo{author}{\bibfnamefont{M.}~\bibnamefont{Chan}},
  \bibinfo{author}{\bibfnamefont{J.}~\bibnamefont{Greeley}},
  \bibinfo{author}{\bibfnamefont{V.}~\bibnamefont{Stamenkovic}},
  \bibnamefont{and} \bibinfo{author}{\bibfnamefont{N.~M.}
  \bibnamefont{Markovic}}, \bibinfo{journal}{Nature materials}
  \textbf{\bibinfo{volume}{11}}, \bibinfo{pages}{550} (\bibinfo{year}{2012}).

\bibitem[{\citenamefont{Bates et~al.}(2015)\citenamefont{Bates, Jia, Doan,
  Liang, and Mukerjee}}]{r8}
\bibinfo{author}{\bibfnamefont{M.~K.} \bibnamefont{Bates}},
  \bibinfo{author}{\bibfnamefont{Q.}~\bibnamefont{Jia}},
  \bibinfo{author}{\bibfnamefont{H.}~\bibnamefont{Doan}},
  \bibinfo{author}{\bibfnamefont{W.}~\bibnamefont{Liang}}, \bibnamefont{and}
  \bibinfo{author}{\bibfnamefont{S.}~\bibnamefont{Mukerjee}},
  \bibinfo{journal}{ACS Catalysis} \textbf{\bibinfo{volume}{6}},
  \bibinfo{pages}{155} (\bibinfo{year}{2015}).

\bibitem[{\citenamefont{Liu et~al.}(2012)\citenamefont{Liu, Cundari, and
  Wilson}}]{r9}
\bibinfo{author}{\bibfnamefont{C.}~\bibnamefont{Liu}},
  \bibinfo{author}{\bibfnamefont{T.~R.} \bibnamefont{Cundari}},
  \bibnamefont{and} \bibinfo{author}{\bibfnamefont{A.~K.}
  \bibnamefont{Wilson}}, \bibinfo{journal}{The Journal of Physical Chemistry C}
  \textbf{\bibinfo{volume}{116}}, \bibinfo{pages}{5681} (\bibinfo{year}{2012}).

\bibitem[{\citenamefont{Aiken and Finke}(1999)}]{aiken1999review}
\bibinfo{author}{\bibfnamefont{J.~D.} \bibnamefont{Aiken}} \bibnamefont{and}
  \bibinfo{author}{\bibfnamefont{R.~G.} \bibnamefont{Finke}},
  \bibinfo{journal}{Journal of Molecular Catalysis A: Chemical}
  \textbf{\bibinfo{volume}{145}}, \bibinfo{pages}{1} (\bibinfo{year}{1999}).

\bibitem[{\citenamefont{Hill and Prosser-McCartha}(1995)}]{hill1995homogeneous}
\bibinfo{author}{\bibfnamefont{C.~L.} \bibnamefont{Hill}} \bibnamefont{and}
  \bibinfo{author}{\bibfnamefont{C.~M.} \bibnamefont{Prosser-McCartha}},
  \bibinfo{journal}{Coordination Chemistry Reviews}
  \textbf{\bibinfo{volume}{143}}, \bibinfo{pages}{407} (\bibinfo{year}{1995}).

\bibitem[{\citenamefont{Albonetti
  et~al.}(1996{\natexlab{a}})\citenamefont{Albonetti, Cavani, and
  Trifiro}}]{r22}
\bibinfo{author}{\bibfnamefont{S.}~\bibnamefont{Albonetti}},
  \bibinfo{author}{\bibfnamefont{F.}~\bibnamefont{Cavani}}, \bibnamefont{and}
  \bibinfo{author}{\bibfnamefont{F.}~\bibnamefont{Trifiro}},
  \bibinfo{journal}{Catalysis Reviews} \textbf{\bibinfo{volume}{38}},
  \bibinfo{pages}{413} (\bibinfo{year}{1996}{\natexlab{a}}).

\bibitem[{\citenamefont{Albonetti
  et~al.}(1996{\natexlab{b}})\citenamefont{Albonetti, Cavani, Trifiro,
  Venturoli, Calestani, Granados, and Fierro}}]{r23}
\bibinfo{author}{\bibfnamefont{S.}~\bibnamefont{Albonetti}},
  \bibinfo{author}{\bibfnamefont{F.}~\bibnamefont{Cavani}},
  \bibinfo{author}{\bibfnamefont{F.}~\bibnamefont{Trifiro}},
  \bibinfo{author}{\bibfnamefont{P.}~\bibnamefont{Venturoli}},
  \bibinfo{author}{\bibfnamefont{G.}~\bibnamefont{Calestani}},
  \bibinfo{author}{\bibfnamefont{M.~L.} \bibnamefont{Granados}},
  \bibnamefont{and} \bibinfo{author}{\bibfnamefont{J.~G.}
  \bibnamefont{Fierro}}, \bibinfo{journal}{Journal of catalysis}
  \textbf{\bibinfo{volume}{160}}, \bibinfo{pages}{52}
  (\bibinfo{year}{1996}{\natexlab{b}}).

\bibitem[{\citenamefont{Ba{\~n}ares}(1999)}]{r24}
\bibinfo{author}{\bibfnamefont{M.~A.} \bibnamefont{Ba{\~n}ares}},
  \bibinfo{journal}{Catalysis Today} \textbf{\bibinfo{volume}{51}},
  \bibinfo{pages}{319} (\bibinfo{year}{1999}).

\bibitem[{\citenamefont{Hodnett}(1985)}]{r25}
\bibinfo{author}{\bibfnamefont{B.}~\bibnamefont{Hodnett}},
  \bibinfo{journal}{Catalysis Reviews Science and Engineering}
  \textbf{\bibinfo{volume}{27}}, \bibinfo{pages}{373} (\bibinfo{year}{1985}).

\bibitem[{\citenamefont{Vining et~al.}(2010)\citenamefont{Vining, Goodrow,
  Strunk, and Bell}}]{r26}
\bibinfo{author}{\bibfnamefont{W.~C.} \bibnamefont{Vining}},
  \bibinfo{author}{\bibfnamefont{A.}~\bibnamefont{Goodrow}},
  \bibinfo{author}{\bibfnamefont{J.}~\bibnamefont{Strunk}}, \bibnamefont{and}
  \bibinfo{author}{\bibfnamefont{A.~T.} \bibnamefont{Bell}},
  \bibinfo{journal}{Journal of Catalysis} \textbf{\bibinfo{volume}{270}},
  \bibinfo{pages}{163} (\bibinfo{year}{2010}).

\bibitem[{\citenamefont{Zhang et~al.}(1995)\citenamefont{Zhang, Desikan, and
  Oyama}}]{r27}
\bibinfo{author}{\bibfnamefont{W.}~\bibnamefont{Zhang}},
  \bibinfo{author}{\bibfnamefont{A.}~\bibnamefont{Desikan}}, \bibnamefont{and}
  \bibinfo{author}{\bibfnamefont{S.~T.} \bibnamefont{Oyama}},
  \bibinfo{journal}{The Journal of Physical Chemistry}
  \textbf{\bibinfo{volume}{99}}, \bibinfo{pages}{14468} (\bibinfo{year}{1995}).

\bibitem[{\citenamefont{Grasselli}(2001)}]{r28}
\bibinfo{author}{\bibfnamefont{R.~K.} \bibnamefont{Grasselli}},
  \bibinfo{journal}{Topics in Catalysis} \textbf{\bibinfo{volume}{15}},
  \bibinfo{pages}{93} (\bibinfo{year}{2001}).

\bibitem[{\citenamefont{Nikolov et~al.}(1991)\citenamefont{Nikolov, Klissurski,
  and Anastasov}}]{r29}
\bibinfo{author}{\bibfnamefont{V.}~\bibnamefont{Nikolov}},
  \bibinfo{author}{\bibfnamefont{D.}~\bibnamefont{Klissurski}},
  \bibnamefont{and}
  \bibinfo{author}{\bibfnamefont{A.}~\bibnamefont{Anastasov}},
  \bibinfo{journal}{Catalysis Reviews} \textbf{\bibinfo{volume}{33}},
  \bibinfo{pages}{319} (\bibinfo{year}{1991}).

\bibitem[{\citenamefont{Heck}(1999)}]{r30}
\bibinfo{author}{\bibfnamefont{R.~M.} \bibnamefont{Heck}},
  \bibinfo{journal}{Catalysis Today} \textbf{\bibinfo{volume}{53}},
  \bibinfo{pages}{519} (\bibinfo{year}{1999}).

\bibitem[{\citenamefont{Li et~al.}(1996)\citenamefont{Li, Huang, and
  Cheng}}]{r31}
\bibinfo{author}{\bibfnamefont{K.-T.} \bibnamefont{Li}},
  \bibinfo{author}{\bibfnamefont{M.-Y.} \bibnamefont{Huang}}, \bibnamefont{and}
  \bibinfo{author}{\bibfnamefont{W.-D.} \bibnamefont{Cheng}},
  \bibinfo{journal}{Industrial \& engineering chemistry research}
  \textbf{\bibinfo{volume}{35}}, \bibinfo{pages}{621} (\bibinfo{year}{1996}).

\bibitem[{\citenamefont{Pal et~al.}(2014)\citenamefont{Pal, Smerieri, Celasco,
  Savio, Vattuone, and Rocca}}]{Pal2014}
\bibinfo{author}{\bibfnamefont{J.}~\bibnamefont{Pal}},
  \bibinfo{author}{\bibfnamefont{M.}~\bibnamefont{Smerieri}},
  \bibinfo{author}{\bibfnamefont{E.}~\bibnamefont{Celasco}},
  \bibinfo{author}{\bibfnamefont{L.}~\bibnamefont{Savio}},
  \bibinfo{author}{\bibfnamefont{L.}~\bibnamefont{Vattuone}}, \bibnamefont{and}
  \bibinfo{author}{\bibfnamefont{M.}~\bibnamefont{Rocca}},
  \bibinfo{journal}{Phys. Rev. Lett.} \textbf{\bibinfo{volume}{112}},
  \bibinfo{pages}{126102} (\bibinfo{year}{2014}),
  \urlprefix\url{https://link.aps.org/doi/10.1103/PhysRevLett.112.126102}.

\bibitem[{\citenamefont{Getsoian et~al.}(2014)\citenamefont{Getsoian, Zhai, and
  Bell}}]{r49}
\bibinfo{author}{\bibfnamefont{A.~?.} \bibnamefont{Getsoian}},
  \bibinfo{author}{\bibfnamefont{Z.}~\bibnamefont{Zhai}}, \bibnamefont{and}
  \bibinfo{author}{\bibfnamefont{A.~T.} \bibnamefont{Bell}},
  \bibinfo{journal}{Journal of the American Chemical Society}
  \textbf{\bibinfo{volume}{136}}, \bibinfo{pages}{13684}
  (\bibinfo{year}{2014}).

\bibitem[{\citenamefont{Bhattacharya et~al.}(2015)\citenamefont{Bhattacharya,
  Sonin, Jumonville, Ghiringhelli, and Marom}}]{SaswataPRBR2015}
\bibinfo{author}{\bibfnamefont{S.}~\bibnamefont{Bhattacharya}},
  \bibinfo{author}{\bibfnamefont{B.~H.} \bibnamefont{Sonin}},
  \bibinfo{author}{\bibfnamefont{C.~J.} \bibnamefont{Jumonville}},
  \bibinfo{author}{\bibfnamefont{L.~M.} \bibnamefont{Ghiringhelli}},
  \bibnamefont{and} \bibinfo{author}{\bibfnamefont{N.}~\bibnamefont{Marom}},
  \bibinfo{journal}{Phys. Rev. B} \textbf{\bibinfo{volume}{91}},
  \bibinfo{pages}{241115} (\bibinfo{year}{2015}),
  \urlprefix\url{https://link.aps.org/doi/10.1103/PhysRevB.91.241115}.

\bibitem[{\citenamefont{Marom et~al.}(2012)\citenamefont{Marom, Kim, and
  Chelikowsky}}]{NoaPRL2012}
\bibinfo{author}{\bibfnamefont{N.}~\bibnamefont{Marom}},
  \bibinfo{author}{\bibfnamefont{M.}~\bibnamefont{Kim}}, \bibnamefont{and}
  \bibinfo{author}{\bibfnamefont{J.~R.} \bibnamefont{Chelikowsky}},
  \bibinfo{journal}{Phys. Rev. Lett.} \textbf{\bibinfo{volume}{108}},
  \bibinfo{pages}{106801} (\bibinfo{year}{2012}),
  \urlprefix\url{https://link.aps.org/doi/10.1103/PhysRevLett.108.106801}.

\bibitem[{\citenamefont{Fukui et~al.}(1952)\citenamefont{Fukui, Yonezawa, and
  Shingu}}]{r43}
\bibinfo{author}{\bibfnamefont{K.}~\bibnamefont{Fukui}},
  \bibinfo{author}{\bibfnamefont{T.}~\bibnamefont{Yonezawa}}, \bibnamefont{and}
  \bibinfo{author}{\bibfnamefont{H.}~\bibnamefont{Shingu}},
  \bibinfo{journal}{The Journal of Chemical Physics}
  \textbf{\bibinfo{volume}{20}}, \bibinfo{pages}{722} (\bibinfo{year}{1952}),
  \eprint{https://doi.org/10.1063/1.1700523},
  \urlprefix\url{https://doi.org/10.1063/1.1700523}.

\bibitem[{\citenamefont{Anslyn and Dougherty}(2006)}]{r44}
\bibinfo{author}{\bibfnamefont{E.~V.} \bibnamefont{Anslyn}} \bibnamefont{and}
  \bibinfo{author}{\bibfnamefont{D.~A.} \bibnamefont{Dougherty}},
  \emph{\bibinfo{title}{Modern physical organic chemistry}}
  (\bibinfo{publisher}{University science books}, \bibinfo{year}{2006}).

\bibitem[{\citenamefont{Khan et~al.}(2016)\citenamefont{Khan, Gupta, Alam, and
  Haider}}]{r50}
\bibinfo{author}{\bibfnamefont{T.~S.} \bibnamefont{Khan}},
  \bibinfo{author}{\bibfnamefont{S.}~\bibnamefont{Gupta}},
  \bibinfo{author}{\bibfnamefont{M.~I.} \bibnamefont{Alam}}, \bibnamefont{and}
  \bibinfo{author}{\bibfnamefont{M.~A.} \bibnamefont{Haider}},
  \bibinfo{journal}{RSC Advances} \textbf{\bibinfo{volume}{6}},
  \bibinfo{pages}{101697} (\bibinfo{year}{2016}).

\bibitem[{\citenamefont{Lopez-Acevedo et~al.}(2010)\citenamefont{Lopez-Acevedo,
  Kacprzak, Akola, and H{\"a}kkinen}}]{r47}
\bibinfo{author}{\bibfnamefont{O.}~\bibnamefont{Lopez-Acevedo}},
  \bibinfo{author}{\bibfnamefont{K.~A.} \bibnamefont{Kacprzak}},
  \bibinfo{author}{\bibfnamefont{J.}~\bibnamefont{Akola}}, \bibnamefont{and}
  \bibinfo{author}{\bibfnamefont{H.}~\bibnamefont{H{\"a}kkinen}},
  \bibinfo{journal}{Nature chemistry} \textbf{\bibinfo{volume}{2}},
  \bibinfo{pages}{329} (\bibinfo{year}{2010}).

\bibitem[{\citenamefont{Fishman et~al.}(2016)\citenamefont{Fishman, Rudshteyn,
  He, Liu, Chaudhuri, Askerka, Haller, Batista, and Pfefferle}}]{r48}
\bibinfo{author}{\bibfnamefont{Z.~S.} \bibnamefont{Fishman}},
  \bibinfo{author}{\bibfnamefont{B.}~\bibnamefont{Rudshteyn}},
  \bibinfo{author}{\bibfnamefont{Y.}~\bibnamefont{He}},
  \bibinfo{author}{\bibfnamefont{B.}~\bibnamefont{Liu}},
  \bibinfo{author}{\bibfnamefont{S.}~\bibnamefont{Chaudhuri}},
  \bibinfo{author}{\bibfnamefont{M.}~\bibnamefont{Askerka}},
  \bibinfo{author}{\bibfnamefont{G.~L.} \bibnamefont{Haller}},
  \bibinfo{author}{\bibfnamefont{V.~S.} \bibnamefont{Batista}},
  \bibnamefont{and} \bibinfo{author}{\bibfnamefont{L.~D.}
  \bibnamefont{Pfefferle}}, \bibinfo{journal}{Journal of the American Chemical
  Society} \textbf{\bibinfo{volume}{138}}, \bibinfo{pages}{10978}
  (\bibinfo{year}{2016}).

\bibitem[{\citenamefont{Raybaud et~al.}(1997)\citenamefont{Raybaud, Hafner,
  Kresse, and Toulhoat}}]{r45}
\bibinfo{author}{\bibfnamefont{P.}~\bibnamefont{Raybaud}},
  \bibinfo{author}{\bibfnamefont{J.}~\bibnamefont{Hafner}},
  \bibinfo{author}{\bibfnamefont{G.}~\bibnamefont{Kresse}}, \bibnamefont{and}
  \bibinfo{author}{\bibfnamefont{H.}~\bibnamefont{Toulhoat}},
  \bibinfo{journal}{Journal of Physics: Condensed Matter}
  \textbf{\bibinfo{volume}{9}}, \bibinfo{pages}{11107} (\bibinfo{year}{1997}).

\bibitem[{\citenamefont{Rodriguez et~al.}(1999)\citenamefont{Rodriguez, Jirsak,
  and Chaturvedi}}]{r51}
\bibinfo{author}{\bibfnamefont{J.~A.} \bibnamefont{Rodriguez}},
  \bibinfo{author}{\bibfnamefont{T.}~\bibnamefont{Jirsak}}, \bibnamefont{and}
  \bibinfo{author}{\bibfnamefont{S.}~\bibnamefont{Chaturvedi}},
  \bibinfo{journal}{The Journal of chemical physics}
  \textbf{\bibinfo{volume}{111}}, \bibinfo{pages}{8077} (\bibinfo{year}{1999}).

\bibitem[{\citenamefont{Blum et~al.}(2009)\citenamefont{Blum, Gehrke, Hanke,
  Havu, Havu, Ren, Reuter, and Scheffler}}]{blum2009ab}
\bibinfo{author}{\bibfnamefont{V.}~\bibnamefont{Blum}},
  \bibinfo{author}{\bibfnamefont{R.}~\bibnamefont{Gehrke}},
  \bibinfo{author}{\bibfnamefont{F.}~\bibnamefont{Hanke}},
  \bibinfo{author}{\bibfnamefont{P.}~\bibnamefont{Havu}},
  \bibinfo{author}{\bibfnamefont{V.}~\bibnamefont{Havu}},
  \bibinfo{author}{\bibfnamefont{X.}~\bibnamefont{Ren}},
  \bibinfo{author}{\bibfnamefont{K.}~\bibnamefont{Reuter}}, \bibnamefont{and}
  \bibinfo{author}{\bibfnamefont{M.}~\bibnamefont{Scheffler}},
  \bibinfo{journal}{Computer Physics Communications}
  \textbf{\bibinfo{volume}{180}}, \bibinfo{pages}{2175} (\bibinfo{year}{2009}).

\bibitem[{\citenamefont{Tkatchenko and
  Scheffler}(2009)}]{tkatchenko2009accurate}
\bibinfo{author}{\bibfnamefont{A.}~\bibnamefont{Tkatchenko}} \bibnamefont{and}
  \bibinfo{author}{\bibfnamefont{M.}~\bibnamefont{Scheffler}},
  \bibinfo{journal}{Physical review letters} \textbf{\bibinfo{volume}{102}},
  \bibinfo{pages}{073005} (\bibinfo{year}{2009}).

\bibitem[{\citenamefont{Perdew et~al.}(1996)\citenamefont{Perdew, Ernzerhof,
  and Burke}}]{perdew1996rationale}
\bibinfo{author}{\bibfnamefont{J.~P.} \bibnamefont{Perdew}},
  \bibinfo{author}{\bibfnamefont{M.}~\bibnamefont{Ernzerhof}},
  \bibnamefont{and} \bibinfo{author}{\bibfnamefont{K.}~\bibnamefont{Burke}},
  \bibinfo{journal}{The Journal of Chemical Physics}
  \textbf{\bibinfo{volume}{105}}, \bibinfo{pages}{9982} (\bibinfo{year}{1996}).

\bibitem[{\citenamefont{Scheffler and Scherz}(1986)}]{scheffler1986resonant}
\bibinfo{author}{\bibfnamefont{M.}~\bibnamefont{Scheffler}} \bibnamefont{and}
  \bibinfo{author}{\bibfnamefont{U.}~\bibnamefont{Scherz}}, in
  \emph{\bibinfo{booktitle}{Materials Science Forum}}
  (\bibinfo{organization}{Trans Tech Publ}, \bibinfo{year}{1986}),
  vol.~\bibinfo{volume}{10}, pp. \bibinfo{pages}{353--358}.

\bibitem[{\citenamefont{Scheffler and
  Dabrowski}(1988)}]{scheffler1988parameter}
\bibinfo{author}{\bibfnamefont{M.}~\bibnamefont{Scheffler}} \bibnamefont{and}
  \bibinfo{author}{\bibfnamefont{J.}~\bibnamefont{Dabrowski}},
  \bibinfo{journal}{Philosophical Magazine A} \textbf{\bibinfo{volume}{58}},
  \bibinfo{pages}{107} (\bibinfo{year}{1988}).

\bibitem[{\citenamefont{Wang et~al.}(1998)\citenamefont{Wang, Weiss,
  Shaikhutdinov, Ritter, Petersen, Wagner, Schl{\"o}gl, and
  Scheffler}}]{wang1998hematite}
\bibinfo{author}{\bibfnamefont{X.-G.} \bibnamefont{Wang}},
  \bibinfo{author}{\bibfnamefont{W.}~\bibnamefont{Weiss}},
  \bibinfo{author}{\bibfnamefont{S.~K.} \bibnamefont{Shaikhutdinov}},
  \bibinfo{author}{\bibfnamefont{M.}~\bibnamefont{Ritter}},
  \bibinfo{author}{\bibfnamefont{M.}~\bibnamefont{Petersen}},
  \bibinfo{author}{\bibfnamefont{F.}~\bibnamefont{Wagner}},
  \bibinfo{author}{\bibfnamefont{R.}~\bibnamefont{Schl{\"o}gl}},
  \bibnamefont{and}
  \bibinfo{author}{\bibfnamefont{M.}~\bibnamefont{Scheffler}},
  \bibinfo{journal}{Physical Review Letters} \textbf{\bibinfo{volume}{81}},
  \bibinfo{pages}{1038} (\bibinfo{year}{1998}).

\bibitem[{\citenamefont{Lee et~al.}(2000)\citenamefont{Lee, Moritz, and
  Scheffler}}]{lee2000gaas}
\bibinfo{author}{\bibfnamefont{S.-H.} \bibnamefont{Lee}},
  \bibinfo{author}{\bibfnamefont{W.}~\bibnamefont{Moritz}}, \bibnamefont{and}
  \bibinfo{author}{\bibfnamefont{M.}~\bibnamefont{Scheffler}},
  \bibinfo{journal}{Physical review letters} \textbf{\bibinfo{volume}{85}},
  \bibinfo{pages}{3890} (\bibinfo{year}{2000}).

\bibitem[{\citenamefont{Reuter and Scheffler}(2003)}]{reuter2003composition}
\bibinfo{author}{\bibfnamefont{K.}~\bibnamefont{Reuter}} \bibnamefont{and}
  \bibinfo{author}{\bibfnamefont{M.}~\bibnamefont{Scheffler}},
  \bibinfo{journal}{Physical Review B} \textbf{\bibinfo{volume}{68}},
  \bibinfo{pages}{045407} (\bibinfo{year}{2003}).

\bibitem[{\citenamefont{Reuter et~al.}(2005)\citenamefont{Reuter, Stampf, and
  Scheffler}}]{reuter2005ab}
\bibinfo{author}{\bibfnamefont{K.}~\bibnamefont{Reuter}},
  \bibinfo{author}{\bibfnamefont{C.}~\bibnamefont{Stampf}}, \bibnamefont{and}
  \bibinfo{author}{\bibfnamefont{M.}~\bibnamefont{Scheffler}}, in
  \emph{\bibinfo{booktitle}{Handbook of Materials Modeling}}
  (\bibinfo{publisher}{Springer}, \bibinfo{year}{2005}), pp.
  \bibinfo{pages}{149--194}.

\bibitem[{\citenamefont{Bhattacharya et~al.}(2017)\citenamefont{Bhattacharya,
  Berger, Reuter, Ghiringhelli, and Levchenko}}]{sasprmr17}
\bibinfo{author}{\bibfnamefont{S.}~\bibnamefont{Bhattacharya}},
  \bibinfo{author}{\bibfnamefont{D.}~\bibnamefont{Berger}},
  \bibinfo{author}{\bibfnamefont{K.}~\bibnamefont{Reuter}},
  \bibinfo{author}{\bibfnamefont{L.~M.} \bibnamefont{Ghiringhelli}},
  \bibnamefont{and} \bibinfo{author}{\bibfnamefont{S.~V.}
  \bibnamefont{Levchenko}}, \bibinfo{journal}{Phys. Rev. Materials}
  \textbf{\bibinfo{volume}{1}}, \bibinfo{pages}{071601} (\bibinfo{year}{2017}),
  \urlprefix\url{https://link.aps.org/doi/10.1103/PhysRevMaterials.1.071601}.

\bibitem[{\citenamefont{Bhattacharya and Bhattacharya}(2015)}]{sasjpcl}
\bibinfo{author}{\bibfnamefont{A.}~\bibnamefont{Bhattacharya}}
  \bibnamefont{and}
  \bibinfo{author}{\bibfnamefont{S.}~\bibnamefont{Bhattacharya}},
  \bibinfo{journal}{The Journal of Physical Chemistry Letters}
  \textbf{\bibinfo{volume}{6}}, \bibinfo{pages}{3726} (\bibinfo{year}{2015}),
  \bibinfo{note}{pMID: 26722747},
  \eprint{https://doi.org/10.1021/acs.jpclett.5b01435},
  \urlprefix\url{https://doi.org/10.1021/acs.jpclett.5b01435}.

\end{thebibliography}

\end{document}